\documentclass[notitlepage,aps,pra,11pt]{revtex4-1}
\usepackage{times,amsmath,amssymb,amsfonts,graphicx}

\newcommand{\ketbra}[2]{\vert #1 \rangle \! \langle #2 \vert}
\begin{document}
\markboth{Claudia Benedetti \and  Matteo G. A. Paris}
{Effective dephasing for a qubit interacting with a 
transverse field}
\title{EFFECTIVE DEPHASING FOR A QUBIT INTERACTING WITH A TRANSVERSE CLASSICAL FIELD}
\author{CLAUDIA BENEDETTI}
\affiliation{Dipartimento di Fisica, Universit\`a degli 
Studi di Milano, I-20133 Milano, Italy
\\claudia.benedetti@unimi.it}
\author{MATTEO G. A. PARIS}
\affiliation{Dipartimento di Fisica, Universit\`a degli Studi di Milano, 
I-20133 Milano, Italy\\
CNISM, UdR Milano, I-20133 Milano, Italy\\
matteo.paris@fisica.unimi.it}
\begin{abstract}
We address the dynamics of a {  qubit interacting with a quasi static
random classical field having both a longitudinal 
and a transverse component and described by a Gaussian
stochastic process.} In particular, we analyze
in details the conditions under which the dynamics may
be effectively approximated by a unitary operation or a pure dephasing without
relaxation.
\end{abstract}
\maketitle
\section{Introduction}
Studying the interaction of a quantum system with its environment
plays a fundamental role in the development of quantum technologies. In
fact, the quantum features of a system, such as the presence of quantum
correlations or superposition of states, are very fragile and may be
destroyed by the action of the environmental noise.  Decoherence may be
induced by classical or quantum noise, i.e. by the interaction with an
environment described classically or quantum-mechanically. The classical 
description is often more realistic to describes environments with 
a very large number of degrees of freedom, or to describe quantum systems 
coupled to a classical fluctuating field. Recently, it has also been shown 
that even certain quantum environments may be described with 
equivalent classical models \cite{helm09,joynt13,sarma13}. Since the environment 
surrounding a quantum system is often composed by a large number of fluctuators, 
it is legitimate to assume a Gaussian statistics for the
noise \cite{tsai04}.  Moreover, the Gaussian approximation is valid even
in the presence of non-Gaussian noise, as far as the coupling with the
environment is weak \cite{galp06,abel08}. 
\par
Among the different classes of open quantum systems, a large attention 
has been paid to qubit systems subject to environmental noise inducing 
a dephasing dynamics \cite{averin04,shnirma,shibata,sarma08,sarma13b,rev13}. 
In this framework, in studying the interaction of a qubit with an external field, it is
often assumed that the typical frequencies of the system are larger than 
the characteristic frequencies of the environment. In these situations 
it is likely that the interaction with the environment induces 
{\em decoherence} through dephasing rather than {\em relaxation} via damping, 
i.e. by inducing transitions between the energy 
levels of the qubit. The effective Hamiltonian 
describing these kind of processes may be thus written as
\begin{equation}
 H(t)=\omega_0\sigma_z+B_z(t)\sigma_z 
\label{Hz}
\end{equation}
where, $\omega_0$ is the natural frequency of the qubit and $B_z(t)$ is
a classical stochastic field with a noise spectrum containing
frequencies that are smaller than $\omega_0$. The overall 
evolution of the system is 
obtained by averaging the unitary evolution governed by the 
Hamiltonian (\ref{Hz}) over the realizations of the stochastic process. 
The resulting map $\rho(t)={\cal E}_t (\rho_0)$ corresponds to
a pure dephasing which, in turn, leads to 
a number of interesting phenomena \cite{rev13}, including the abrupt
vanishing of entanglement 
(the so-called entanglement sudden-death \cite{esd1,esd2,esd3}) and the 
sudden transition between classical and quantum decoherence
\cite{sd1,sd2}. Pure dephasing has been also used to 
describe the dynamics of qubit systems in colored environments
\cite{ben12,ben13} and to quantify their non-Markovian character
\cite{ben13a}.
\par
In this paper we do not assume the Hamiltonian in Eq. (\ref{Hz}), 
and address the dynamics of a qubit interacting
with a Gaussian field with both a longitudinal and a
transverse component and with a broad spectrum, possibly
including the natural frequency $\omega_0$ of the qubit.
In particular, we are interested in analyzing the conditions
under which the dynamics may be effectively approximated by a unitary
operation or a pure dephasing without relaxation.
{  Addressing the problem for a generic transverse stochastic field 
is a challenging task \cite{cum1,cum2} since a high order
cumulant expansion is involved.
We thus restrict attention to the quasi static regime, where 
the dynamics of the external field is assumed to be slow, and
discuss in some details the conditions to obtain an effective 
dephasing in this regime.}
\par
The paper is structured as follows: In Section \ref{s:qb}
we describe the dynamics of a qubit interacting with an external 
random classical field having nonzero longitudinal and 
transverse components. In Section
\ref{s:tr} we assume a pure transverse field and analyze the conditions under 
which its effects may be neglected, i.e. the dynamics may be effectively 
approximated by a unitary operation or a dephasing, whereas in Section \ref{s:fl} we
consider both components and again analyze the regimes where the dynamics 
corresponds to dephasing without relaxation.
Section \ref{s:out} closes the paper with some concluding remarks.
\section{Qubit interacting with a classical random field}
\label{s:qb}
Let us consider a two level system interacting with an external 
fluctuacting field $\vec{B}$, having both a longitudinal and a 
transverse component, denoted by $B_z(t)$ and $B_x(t)$ respectively.
The system Hamiltonian is given by:
\begin{align}
 H(t)&=\omega_0\sigma_z+B_x(t)\sigma_x+B_z(t)\sigma_z \label{hamiltonian},
\end{align}
where $\omega_0$ is the qubit energy and the $\sigma_i$ are Pauli matrices.
Our purpose is to study under which conditions the dynamics governed by
the Hamiltonian \eqref{hamiltonian}, and by the average over the
stochastic processes $B_z(t)$ and $B_x(t)$, may be described by a
dephasing map, such that the added term $B_x(t)\sigma_x$ does not affect
the population of the qubit.
The time-dependent coefficients $B_i(t)$ describe stationary Gaussian stochastic
processes with zero mean and covariance $K(t,t')\equiv 
K(t-t')$, in formula 
\begin{align}
 [B_i(t)]_{B_i}&=0\nonumber\\
 [B_i(t)B_i(t')]_{B_i}&=K_i(t-t')\qquad i=x,z
\end{align}
where the symbol $[\cdot]_{B_i}$ denotes the average over the 
process $B_i(t)$. A Gaussian process is a process
which can be fully described by its second-order statistics. The 
characteristic function is given by \cite{puri}
\begin{equation}
 \left[ \exp\left(i\int_{t_0}^t\!\!\! ds\, J(s) B_i(s)\right)\right]_{B_i}= 
 \hbox{exp}\left(-\frac{1}{2}
 \int_{t_0}^t 
 \int_{t_0}^t \!\!ds\,ds'\,J(s) K_i(s-s')J(s')\right).
 \label{gauSt}
\end{equation}
{  Upon assuming $t_0=0$, the evolution operator is expressed as:
\begin{align}
U(t,\omega_0)& =\exp\left\{-i\, {\cal T}\int_0^t\!\!ds\, H(s)\right\}
\notag \\ &\simeq
\exp\left\{-i \left[\,\omega_0 t\, \sigma_z +\varphi_x(t)\,
\sigma_x+\varphi_z(t)\,\sigma_z\right]\right\}\label{uu}
\end{align}
where ${\cal T}$ denotes time ordering operator and we have introduced the noise phases 
$$\varphi_i(t)=\int_0^t\!\! ds\, B_i(s)\,.$$
The second equality in Eq. (\ref{uu}) is only approximated and is valid upon truncating
the Dyson series at the first order, i.e. assuming that we are in the 
quasi static regime such that the two-time commutator $[H(t_1),H(t_2)]$ is 
negligible. If the external field is exactly static, i.e. it is random
but it does not change in time, the phases are given by $\varphi_i(t)=
B_i(s)\,t$ while in the quasi static regime they encompass the effects
of the (slow) dynamics of the external field.}
Because of the Gaussian nature of the considered process, the average of 
any functional of the noise phase $g[\varphi(t)]$ may be written as the the average over
the process $\varphi(t)$ with a Gaussian probability distribution: 
\begin{align}
 [g(\varphi_i)]_{B_i}&=
 \frac{1}{\sqrt{2\pi\beta_i(t)}}
 \int\!\! d\varphi_i\, g(\varphi_i)\,
 \exp\left\{-\frac{\varphi_i^2}{2\beta_i(t)}\right\}
\end{align}
where we omitted the explicit dependency of $\varphi$ on time, and
the variance function $\beta(t)$ is defined as:
\begin{align}
\beta_i(t)=\int_0^t\int_0^t \!\!ds\,ds'\, K_i(s-s').
 \label{beta}
\end{align}
The evolution operator may be decomposed into the 
Pauli basis,  $U(t,\omega_0)=\frac{1}{2}\sum_{j=0}^4 
\hbox{Tr}[U(t,\omega_0)\sigma_j]\sigma_j$,
with $\sigma_0$ corresponding to the identity matrix $\mathbb{I}$,
and can thus be expressed as:
\begin{align}
U(t,\omega_0)=f_I(t,\omega_0)\,\mathbb{I}+i\,f_x(t,\omega_0)\,\sigma_x
+i\,f_z(t,\omega_0)\,\sigma_z,
\end{align}
where
\begin{align}
 f_I(t,\omega_0)&=\cos\left[\sqrt{\varphi_x^2+(\varphi_z+\omega_0 t)^2}\right]\\
f_x(t,\omega_0)&=-\frac{\varphi_x\sin\left[
\sqrt{\varphi_x^2+(\varphi_z+\omega_0 t)^2}
\right]}{\sqrt{\varphi_x^2+(\varphi_z+\omega_0 t)^2}}
\label{fx}\\
f_z(t,\omega_0)&=-\frac{(\varphi_z+\omega_0 t)\sin
\left[\sqrt{\varphi_x^2+(\varphi_z+\omega_0 t)^2}
\right]}{\sqrt{\varphi_x^2+(\varphi_z+\omega_0 t)^2}}\,.
\end{align}
The qubit density matrix is then evaluated as the average of the evolved density matrix
over the stochastic processes $\vec{B}=\{B_x,B_z\}$:
\begin{align}
\rho(t)=\left[U(t,\omega_0)\rho_0U^{\dagger}(t,\omega_0)\right]_{\vec{B}}
\label{rhot}
\end{align}
where $\rho_0=\sum_{j,k=1}^2\rho_{jk}\ketbra{j}{k}$ is the initial
density operator. Since the average of any odd terms 
in $\varphi_x$ and $\varphi_z$ 
in Eq. \eqref{rhot} vanishes, we have
\begin{align}
 &\rho(t)=
 \Bigg[f_I^2\,\rho_0+f_x^2\,\sigma_x\rho_0\sigma_x+
 f_z^2\,\sigma_z\rho_0\sigma_z+i\,f_If_z\,[\sigma_z,\rho_0]\,\Bigg]_{\vec{B}}
 \label{evol}
\end{align}
where we omitted the dependency of the $f$ functions on $t$ and
$\omega_0$. After performing the average in Eq. \eqref{evol}, the 
evolved density matrix may be rewritten as:
\begin{align}
 \rho(t)=A_I\,\rho_0+A_x\, \sigma_x\rho_0\sigma_x+A_z\,\sigma_z\rho_0\sigma_z
 +i A_{Iz}\,[\sigma_z,\rho_0]
 \label{rho_A}
\end{align}
where: 
\begin{align}
 A_i&=A_i(t,\omega_0)=\Big[f_i(t,\omega_0)^2\Big]_{\vec{B}}\qquad i=I,x,z\label{Ai}\\
 A_{Iz}&=A_{Iz}(t,\omega_0)=\Big[f_{I}(t,\omega_0)f_{z}(t,\omega_0)\Big]_{\vec{B}}
\end{align}
and the condition $A_I+A_x+A_z=1$ must be satisfied to preserve
unitarity. Upon writing explicitly the density matrix 
\begin{equation}
 \rho(t)=\left(\begin{array}{cc}
 (A_I+A_z)\rho_{11}+A_x\rho_{22}&(A_I+2i\;A_{Iz}-A_z)\rho_{12}+A_x\rho_{21}\\
 A_x\rho_{21}+(A_I-2i\;A_{Iz}-A_z)\rho_{21}&A_x\rho_{11}+(A_I+A_z)\rho_{22}
\end{array}
 \right)\label{rho}
\end{equation}
one immediately sees that whenever $A_x$ is vanishing or may be
neglected, the Hamiltonian \eqref{hamiltonian} leads to a dephasing map,
with a complex dephasing coefficient. In the next Section, we analyze
whether this is true also in other conditions.
\subsection{Interaction with a pure transverse field}
\label{s:tr}
\begin{figure}[h!]
\centering
\includegraphics[width=0.32\textwidth]{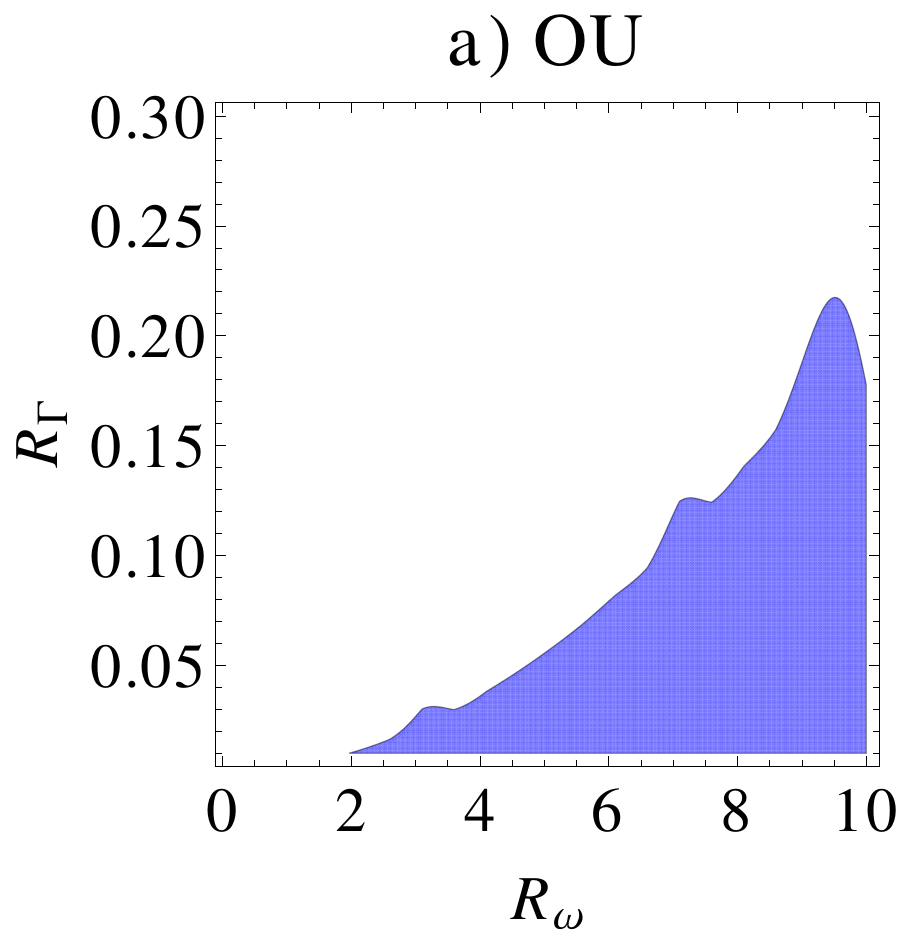}
\includegraphics[width=0.32\textwidth]{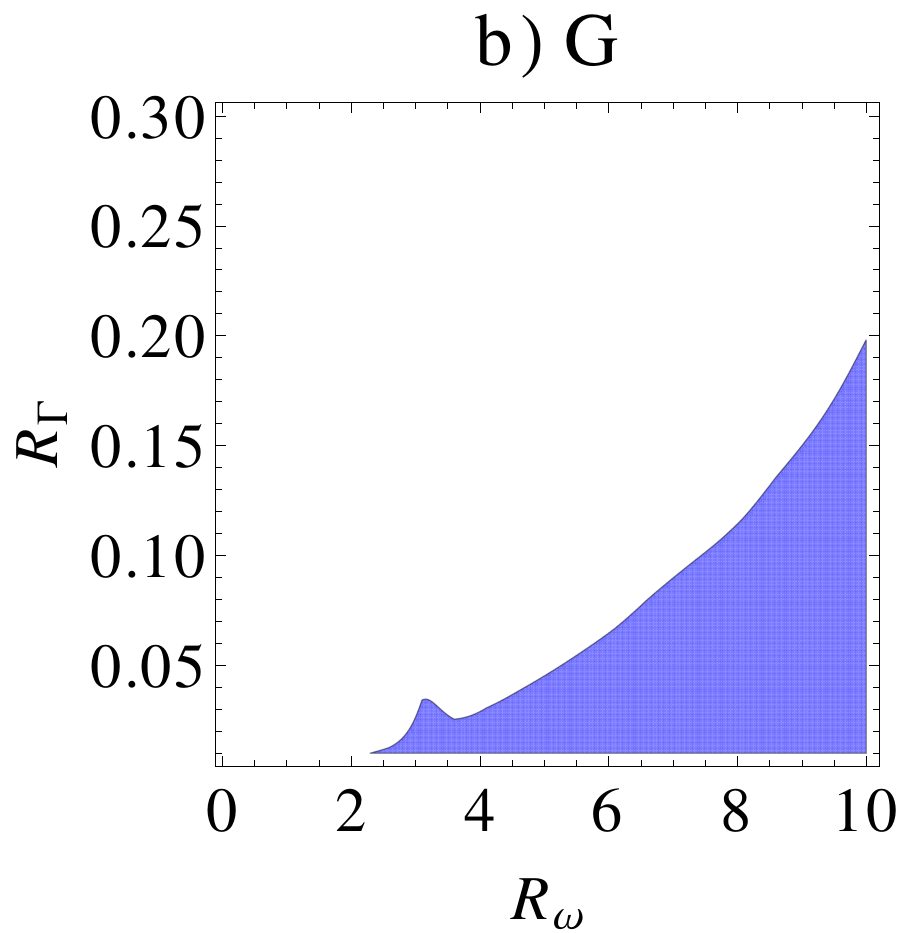}
\includegraphics[width=0.32\textwidth]{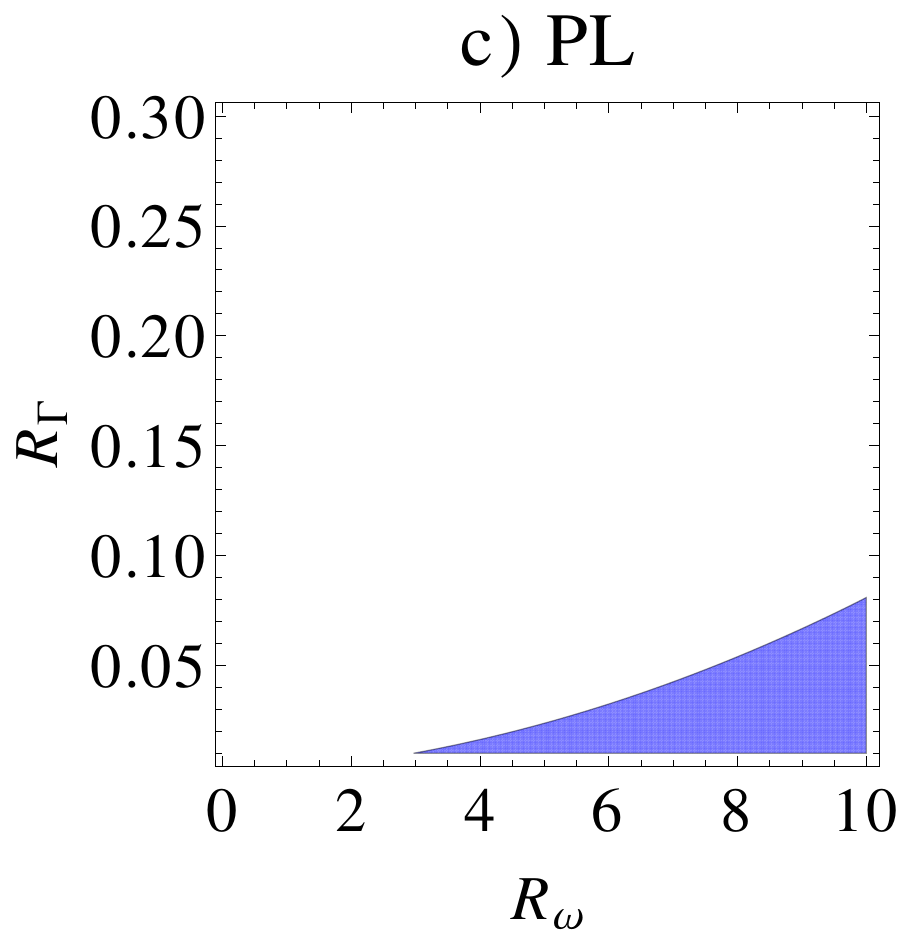}
\caption{Region where the coefficient $\max_\tau 
|A_x(\tau,R_{\omega_0},R_{\Gamma})|<10^{-3}$
for three different processes characterized by a) exponential
b) Gaussian and c) power law ($\alpha=4$) autocorrelation function}
\label{f1}
\end{figure}

In order to gain insight into the dynamics of the system let us first
consider the case of zero longitudinal field $B_z(t)=0$ and look for the
conditions under which the effects of the transverse field may be
neglected or subsumed by a dephasing. 
We set $\varphi_x=\varphi$ and evaluate $A_x(t)$ from Eq. 
\eqref{Ai}, which now reads
\begin{align}
 A_x(t,\omega_0,\beta)&=
 \frac{1}{\sqrt{2\pi\beta(t)}}\int_{-\infty}^{\infty} 
 \!\!\!\!d\varphi\; \varphi^2\,\frac{\sin^2\left[\sqrt{\varphi^2+(\omega_0 t)^2}\right]}
 {\varphi^2+(\omega_0
 t)^2}\,\exp\left(-\frac{\varphi^2}{2\beta(t)}\right)\,,\label{axx}
\end{align}
where the exact functional form of the variance $\beta(t)$ depend on the
specific features of the process $B_x(t)$.
Upon inspecting Eq. (\ref{axx}) one sees that $A_x(t,\omega_0,\beta)$
vanishes whenever $\omega_0 t\gg1$ or $\beta(t)\ll1$. The first condition 
corresponds to the assumption of a large qubit frequency
(outside the spectrum of the noise), whereas the second one
$\beta\ll1$ is related to the specific properties of the stochastic 
process describing the noise. In order to better understand the effects 
of the transverse field, 
we now  evaluate the function $A_x(t,\omega_0,\beta)$ from Eq. \eqref{axx} for 
three classical Gaussian processes with Ornstein-Uhlenbeck (OU), Gaussian (G) 
and power-law (PL) autocorrelation function, i.e. 
\begin{align}
 K_{OU}(t-t', \gamma,\Gamma)&=\frac12\, \Gamma\gamma \,e^{-\gamma|t-t'|}\quad\\
 K_{G}(t-t',\gamma,\Gamma)&=\frac{1}{\sqrt{\pi}}\,\Gamma\gamma\,e^{-\gamma^2(t-t')^2}\\
 K_{PL}(t-t',\gamma,\Gamma,\alpha)&=\frac{1}{2}\,(\alpha-1)\,
 \gamma\Gamma\,\frac{1}{\big(\gamma|t-t'|+1\big)^{\alpha}}\label{plc}
\end{align}
which, by Eq. \eqref{beta}, give:
\begin{align}
 \beta_{OU}(\tau, R_{\Gamma})
&=R_{\Gamma}\left(\tau-1+e^{-\tau}\right)\label{beta_1}
\equiv R_\Gamma\, g_{OU} (\tau)\,,\\
 \beta_{G}(\tau, R_{\Gamma})&=\frac{R_{\Gamma}}{\sqrt{\pi}}
\left[e^{- \tau^2}-1+\sqrt{\pi}\,\tau\, \hbox{Erf}(\tau)\right]
\equiv R_\Gamma\, g_{G} (\tau)\,,\\
 \beta_{PL}(\tau, R_{\Gamma},\alpha)&=R_{\Gamma}\,
\frac{(1-\tau)^2+(1+\tau)^{\alpha}[\tau(\alpha-2)-1]}{(1+\tau)^{\alpha}(\alpha-2)}
\equiv R_\Gamma\, g_{PL} (\tau)]\,,
\label{beta_3}
\end{align}
where $\Gamma$ and $\gamma$ are the {\em damping} and the {\em memory} parameters
of the processes, $\tau=\gamma t$ denotes the rescaled dimensionless time, 
$R_{\Gamma}=\frac{\Gamma}{\gamma}$, $\alpha>2$
is a real number and Erf(x) is the error function. {  The (quasi) 
static limit is obtained for vanishing $\gamma$ keeping 
$\Gamma\gamma$ finite.} 
The $g_x(\tau)$'s are
functions of the sole rescaled time, $x=OU,G,PL$.
We have numerically evaluated the integral in Eq.
\eqref{axx} for the three different process as a function of 
rescaled time $\tau$ and the two ratios $R_{\omega}=\omega_0/\gamma$ and $R_{\Gamma}$.
In particular, we want to see when $A_x(\tau,R_{\omega},R_{\Gamma})$ 
is negligible, as a function of the parameters $R_{\omega}$ and
$R_{\Gamma}$, and to this aim, we have maximized the function over the time $\tau$ 
and determined where the maximum is smaller than a given threshold.
In Fig \ref{f1} we show the region in the $R_\omega$--$R_\Gamma$ plane 
where $\max_\tau  |A_x(\tau,R_{\omega_0},R_{\Gamma})|<10^{-3}$
for the three different processes. As it is apparent from the plots, 
the coefficient is negligible if $R_{\omega}\gg1$ and/or $R_{\Gamma}\ll1$,
with the specific ranges depending on the chosen process.
\par
In Fig. \ref{f1}c, we have shown results for a powerlaw process 
with $\alpha=4$. This is a good representative of the 
family \eqref{plc}, since different values of the 
parameter lead to the same conditions for an 
effective dephasing. The behaviour emerging from Fig. \ref{f1} is in agreement 
with the qualitative considerations made above and with the fact that
the condition $\beta\ll1$ is equivalent to $R_{\Gamma}\ll 1$. 
\par
Since we assumed Gaussian processes with zero mean, we can Taylor-expand
the function $f_x(t,\omega_0)^2$ around $\varphi=0$. By dropping the 
expansion at the second order we may analytically compute the integral 
\eqref{Ai} and obtain:
\begin{align}
\tilde{A}_x(t,\omega_0) &\simeq\beta(t)\frac{\sin^2\omega_0t}{(\omega_0t)^2}.
\label{Ax}
\end{align}
>From Eq. \eqref{Ax} we immediately see that the coefficient $A_x(t,\omega_0)$ 
vanishes for vanishing $\beta(t)$ or for $R_\omega \tau \equiv \omega_0t\gg1$.
This is in agreement with the numerical results and shows
that a second order expansion is sufficient to capture the 
two regimes where the effects of the transverse field on the 
populations may be neglected. In order to gain more insight on the possible
differences between the two regimes we expand, up to 
second order in $\varphi$, also the other $f$ functions, arriving at 
\begin{align}
 \tilde{A}_I(t,\omega_0)&\simeq\cos^2\omega_0 t 
 - \beta(t) \frac{\sin 2\omega_0t}{2\omega_0t}\\
 \tilde{A}_z(t,\omega_0)&\simeq\sin^2\omega_0 t
 - \beta(t) \left (
 \frac{\sin^2 \omega_0t}{(\omega_0t)^2} -
 \frac{\sin 2\omega_0t}{2\omega_0t}\right) \\
 \tilde{A}_{Iz}(t,\omega_0)&\simeq-\frac{1}{2}\sin2\omega_0 t 
 - \beta(t) \left (
 \frac{\cos 2\omega_0t}{2\omega_0t}-
 \frac{\sin 2\omega_0t}{4(\omega_0t)^2} 
 \right)\,.
\end{align}
In turn, the coefficient in the off-diagonal 
elements of the density matrix reads as follows
\begin{align}\label{offd}
A_I+2i\,A_{Iz}-A_z 
&\simeq 
e^{-2 i \omega_0 t} + \frac{\beta(t)}{2(\omega_0 t)^2}& R_\omega \gg 1
\\
&\simeq 
e^{-2 i \omega_0 t} \left[
1-
\frac{\beta(t)}{2(\omega_0 t)^2}
-i\,\frac{\beta(t)}{\omega_0 t}
\right]
+ \frac{\beta(t)}{2(\omega_0 t)^2}
& R_\Gamma \ll 1\,.\label{offd1}
\end{align}
The above expressions, together with Eq. (\ref{Ax}) which is valid in
both the limiting cases, illustrate the qualitative differences 
between the two regimes: for $R_\omega\gg1$ the leading terms in Eqs.
(\ref{offd}) and (\ref{Ax}) are the same, meaning that either relaxation
occurs or the dynamics is unitary, whereas for $R_\Gamma\ll1$ the
multiplicative term in Eq. (\ref{offd1}) reveals that the effective
dynamics of the qubit corresponds to a dephasing. 
{  The expressions above correspond to situations where the effective 
dynamics is valid at all times. More generally, it may happen
that the weaker conditions $R_{\omega}\tau \gg1$ and 
$R_{\Gamma}\,g_x(\tau)\ll1$ are satisfied for up to some values of 
$\tau$, corresponding to regimes where the effective dynamics appears 
only for a finite interaction time.} 
\subsection{Effective dephasing in the general case}
\label{s:fl}
We now consider the complete Hamiltonian \eqref{hamiltonian},
with the longitudinal term $B_z(t)\neq0$.  The coefficient $A_x$, in this
case, takes the form:
\begin{align}
 &A_x(\beta_x,\beta_z,t,\omega_0)=\nonumber\\
 &\frac{1}{2\pi\sqrt{\beta_x(t)\beta_z(t)}}\int\!\!\int\!\! 
 d\varphi_x\,d\varphi_z\,
 \exp\left(-\frac{\varphi_x^2}{2\beta_x(t)}-\frac{\varphi_z^2}{2\beta_z(t)}\right)\,
 f_x^2(\varphi_x,\varphi_z,t,\omega_0)
 \end{align}
 where we explicitly wrote the dependency on the $\beta$ functions.
 Following the line of reasoning of the previous section, we 
 expand the function $f_x$ in Eq. \eqref{fx} around
$\varphi_x=0$ and $\varphi_z=0$, and we drop the expansion at the 
second order. Inserting this expansion in Eq. \eqref{Ai}, we are able
to write the analytical expression for $\tilde{A}_x$:
\begin{align}
\tilde{A}_x(\beta_x,\beta_z,t,\omega_0)=& \beta_x (t)\frac{\sin^2(2\omega_0t)}{\omega_0^2t^2}
+ \notag \\
& \frac{\beta_x(t)\beta_z(t)}{2 (\omega_0 t)^4} \Big[
3 + (2 \omega_0^2 t^2 -3)\cos 2 \omega_0 t - 4 \omega_0 t \sin 2 \omega_0 t
\Big]
\end{align}
Upon expanding to the second order all the terms we may write the
analytical expression of the evolved density matrix, where the off-diagonal 
coefficient $K=A_I+2i\,A_{Iz}-A_z$ is given by:
\begin{align}
K\simeq&\,e^{-2i \omega_0 t}[1-2\beta_z(t)]+
\frac{\beta_x(t)}{2\omega_0^2t^2}&\quad R_{\omega}\gg1\label{r1}\\
 \simeq&\, e^{-2i\,\omega_0t}\left[ 1-2\beta_z (t) -
 \frac{\beta_x(t)}{2 \omega_0^2 t^2} - i \frac{\beta_x(t)}{\omega_0t}\right]	
 +\frac{\beta_x(t)}{2\omega_0^2t^2} &\quad R_{\Gamma}\ll 1\,.  
	       \label{r2}
\end{align}
Looking at Eq. (\ref{r1}) and Eq. \eqref{r2}, one sees that when 
$R_\omega \gg1$ one may just neglect
the effects of the transverse field, whereas for $R_\Gamma \ll 1$ one
has an additional effective term in the coefficient $K$.
{ {As for the previous case, the effective dynamics emerges 
if the above conditions are valid at all times.
More general regimes can be written as  $R_{\omega}\tau \gg1$ and $R_{\Gamma}\,g_x(\tau)\ll1$. }}
\section{Conclusions}
\label{s:out}
The effect of classical noise on a qubit system may be described as the
interaction with a random field. In this paper we analyzed, in the quasi
static regime, the conditions under which a
general dynamics, including interaction with a transverse field, 
may be approximated by an effective dephasing, without
changes in the populations. In particular, we
studied the time evolution of a qubit subject to a transverse and
longitudinal field.  We found  that the properties of the
stochastic processes analyzed, i.e. the autocorrelation function, play a
role, through the variance function $\beta(t)$.  Whenever this
function is small, the dynamics can be described as a dephasing.
Moreover, we recovered the known condition of large system's
energy, {  $\omega_0t\gg1$}, which prevents jumps between the qubit
levels.  If these assumptions do not hold, the general dynamics
is not a dephasing and relaxation phenomena may occur, with changes in
the qubit populations as described by Eq. \eqref{rho}.
\section*{Acknowledgments}
This work has been partially supported by MIUR (FIRB LiCHIS-RBFR10YQ3H)
and by the Finnish Cultural Foundation (Science Workshop on
Entanglement). The authors thank {\L}ukasz Cywinski, P. Bordone, 
F. Buscemi, F. Caruso, A. D'Arrigo, S. Maniscalco and E.
Paladino for discussions and suggestions, and the University of Modena and
Reggio-Emilia for hospitality.

\end{document}